\documentclass[aps, prb, reprint, superscriptaddress, preprintnumbers,amsmath,amssymb,twocolumn,floatfix]{revtex4-2}
\usepackage{amssymb}
\usepackage{amsmath}
\usepackage{bm}
\usepackage[dvips]{graphicx}
\usepackage{dcolumn}
\usepackage{longtable}
\usepackage{color}
\usepackage[version = 4]{mhchem}
\usepackage{comment}

\newcommand{\subm}[1]{_{\mathrm{#1}}}  

\newcommand{\CRA}{CeRh$_2$As$_2$~}

\begin{document}
\title{Incommensurate modulation with $\bm{Q}=\bm{0}$ A-type Antiferromagnetic Order in \CRA revealed by NQR studies}

\author{Shiki \surname{Ogata}}
\email{ogata.s.1029@m.isct.ac.jp}
\affiliation{Department of Physics, Kyoto University, Kyoto 606-8502, Japan}
\author{Shunsaku \surname{Kitagawa}}
\affiliation{Department of Physics, Kyoto University, Kyoto 606-8502, Japan}
\author{Kenji \surname{Ishida}}
\affiliation{Department of Physics, Kyoto University, Kyoto 606-8502, Japan}

\author{Manuel \surname{Brando}}
\affiliation{Max Planck Institute for Chemical Physics of Solids, D-01187 Dresden, Germany}

\author{Elena \surname{Hassinger}}
\affiliation{Institute for Quantum Materials and Technology, Karlsruhe Institute of Technology, 76131 Karlsruhe, Germany}
\affiliation{Max Planck Institute for Chemical Physics of Solids, D-01187 Dresden, Germany}

\author{Christoph \surname{Geibel}}
\affiliation{Max Planck Institute for Chemical Physics of Solids, D-01187 Dresden, Germany}
\author{Seunghyun \surname{Khim}}
\affiliation{Max Planck Institute for Chemical Physics of Solids, D-01187 Dresden, Germany}
\date{\today}

\begin{abstract}
We performed $^{75}$As nuclear quadrupole resonance (NQR) and nuclear
magnetic resonance (NMR) measurements on a higher-quality single-crystalline
CeRh$_2$As$_2$, a heavy-fermion superconductor exhibiting multiple
superconducting (SC) phases under magnetic fields along the $c$ axis. This
SC multiphase is believed to originate from staggered Rashba spin–orbit
coupling associated with locally broken inversion symmetry. In addition to superconductivity, \CRA exhibits phase I below $T_0\sim0.5$ K and an antiferromagnetic (AFM) state below $T\subm{N}\sim 0.25$ K in the early-stage samples. In the higher-quality sample, the AFM transition becomes more pronounced, and $T\subm{N}$ increases to nearly coincide with $T\subm{SC}$. The NQR spectra at the As(1) site imply an internal field with an incommensurate distribution, indicating a two-dimensional incommensurate modulation of the magnetic structure superimposed on a $\bm{Q}=\bm{0}$ A-type AFM component. Moreover, a pronounced decrease in the NQR intensity at $T_0$ well-above $T\subm{N}$ and an abrupt increase in the internal field at $T\subm{N}$ suggest the emergence of a slowly fluctuating AFM order at $T_0$ which becomes static at $T\subm{N}$.
\end{abstract} 

\maketitle

%
\CRA is a heavy-fermion superconductor ($T\subm{SC}\sim0.37$ K) discovered in 2021 \cite{hakken}. The crystal structure of \CRA is of tetragonal CaBe$_2$Ge$_2$-type local noncentrosymmetric structure with the space group $P4/nmm$ (No.~129, $D^{7}_{4h})$. There are two crystallographically inequivalent As and Rh sites; As(1) [Rh(1)] is tetrahedrally coordinated by Rh(2) [As(2)], as shown in Fig.~1(a). \CRA exhibits a SC multiphase under a $c$-axis magnetic field with a SC-SC transition at $\mu_{0}H^{*}\sim 4$ T \cite{hakken}. The SC multiphase is attributed to even-parity and odd-parity SC states due to a staggered Rashba spin-orbit coupling at the Ce site, arising from the local noncentrosymmetric structure \cite{T.Yoshida, Sigrist_PDW_2014, hakken}. Below $T_{0}\sim 0.5$ K, \CRA exhibits a non-trivial ordered phase (phase I) \cite{QDW, QDW_theta, QDW_new, QDW_Los, QDW_Khanenko}. Two possibilities for the order parameter in the phase I have been discussed: a non-magnetic electric quadrupole density wave state arising from hybridization of the crystal electric field (CEF) levels \cite{QDW, Kondo_CEFhybridization_Xray}, or a weak magnetic order \cite{muSR, microHall, T0phasediagram_theory}. Under in-plane magnetic fields, $T_0$ is enhanced, and there exhibits a transition to another phase (phase II) above 9 T \cite{QDW}. Recently, sample quality has improved, resulting in higher $T\subm{SC}$ and $T_0$ compared to the early-stage samples and more pronounced anomalies at $T_0$ in various measurements \cite{QDW_new, QDW_Khanenko}.\par
We reported in previous nuclear quadrupole resonance (NQR) measurement that \CRA also exhibits an AFM state below $T\subm{N}\sim0.25$ K, which is lower than $T\subm{SC}\sim0.37$ K \cite{Kibune}. The anomaly below $T\subm{SC}$ was also reported from the specific-heat measurement \cite{Poland_AFM_1storder}. Significant broadening of the NQR spectrum was observed at the As(2) site while not observed at the As(1) site, suggesting the internal field of the AFM order and its cancellation at the As(1) site \cite{Kibune}. Comparison between nuclear magnetic resonance (NMR) spectra along the $c$-axis and these along the [110] directions revealed that the internal field at the As(2) site is oriented to the $c$ axis \cite{Ogata_ab_old}. These results suggest the $\bm{Q}=\bm{0}$ A-type AFM order with magnetic moment aligned to the $c$ axis, as shown in Fig. 1(a) \cite{Ogata_ab_old,AFMandT0andSC_theory}. In the $c$-axis field, the AFM order disappears around 4 T, suggesting a correlation with the SC multiphase \cite{Ogata_c_old}. In general, the AFM order below $T\subm{SC}$ is unusual and distinct from the coexistence of magnetism and superconductivity usually observed in heavy-fermion and Fe-based superconductors \cite{CeRhIn5_pressure,CeRhIn5_PandH, BaFe2As2_Codope_Pressure}. Therefore, the AFM order in \CRA possibly exhibits a unique property distinct from the magnetism that drives superconductivity in most of heavy-fermion superconductors. However, it is also possible that the magnetism starts at $T\subm{SC}$ or $T_0$, and cannot be detected due to the measurement constraints. In fact, muon spin rotation/relaxation ($\mu$SR) measurements \cite{muSR} and micro-Hall probe measurement \cite{microHall} using higher-quality samples suggested that magnetic order emerges at $T_0$. Since our previous NQR/NMR measurements were carried out on the early-stage sample, we need to investigate the magnetism in more detail using a higher-quality sample.
\par
The higher-quality single crystals of \CRA with a typical size of 1.5 $\times$ 1.0 $\times$ 0.5 mm$^3$ were grown by the modified Bi-flux method \cite{QDW_new}. $T\subm{SC} = $ 0.37 K was determined by the onset temperature of the SC diagmagnetic signal from the ac susceptibility measurement using an NQR coil. NQR/NMR measurements were performed using a $^3$He-$^4$He dilution refrigerator, in which the sample was immersed in the $^3$He-$^4$He mixture to reduce radio-frequency (RF) heating during measurements. For the NMR measurements, we used a split SC magnet that generates a horizontal field and combined it with a single-axis rotator to apply a magnetic field parallel or perpendicular to the $c$ axis. The magnetic field was calibrated using $^{63}$Cu- and $^{65}$Cu-NMR signals from the NMR coil (nuclear gyromagnetic ratio $^{63}\gamma /2\pi = 11.289$ MHz/T and $^{65}\gamma /2\pi = 12.093$ MHz/T, respectively). We experimentally confirmed superconductivity just after the RF pulses using a technique reported in previous studies \cite{Heat-up_SRO,UTe2_Knight_shift}. The $^{75}$As-NQR and NMR spectra (nuclear spin $I$ = 3/2, $^{75}\gamma/2\pi = 7.290$ MHz/T) were obtained as a function of frequency at zero field and fixed magnetic fields, respectively. The site assignment of the NMR peaks was described in the previous paper \cite{Kibune}.\par
%
The higher quality of the sample was verified from the following results. Figures 1(b) and 1(c) show the $^{75}$As-NQR spectra compared to those measured in the early-stage sample \cite{Kibune, Kibune_normal}. The NQR frequency distributions $\Delta \nu\subm{Q}/\nu\subm{Q}$, which reflect sample quality, are $\Delta \nu\subm{Q}/\nu\subm{Q} \sim 0.13 \%$ at the As(1) site and $\Delta \nu\subm{Q}/\nu\subm{Q} \sim 0.39 \%$ at the As(2) site, respectively. Here, $\Delta\nu\subm{Q}$ are estimated from the full width at half maximum (FWHM) of the NQR spectrum. $\Delta \nu\subm{Q}/\nu\subm{Q}$ were less than half those of the early-stage sample ($\sim$ 0.27 \% at the As(1) site and $\sim$ 1.0 \% at the As(2) site), indicating improved sample quality from a microscopic point of view. The higher-frequency shoulder structure observed in the As(2) NQR spectrum in the early-stage sample is less pronounced in the higher-quality sample (only small peak at roughly 10.9 MHz). This suggests that the shoulder structure is due to structural disorder. Figure 1(d) shows the SC diamagnetic signal from ac susceptibility measurements. Although the onset temperature of the SC transition is slightly lower, the transition is much sharper than that in the early-stage sample, indicating a more uniform SC transition.
\begin{figure*}[htbp!]
   \begin{center}
   \includegraphics[width=0.95\linewidth]{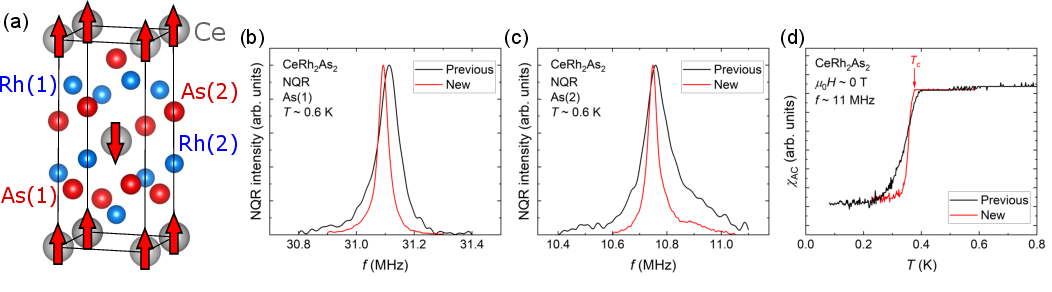}
   \end{center}
   \caption{(a) Magnetic structure of \CRA \cite{VESTA}. (b)(c) The normal state $^{75}$As-NQR spectra of the As(1) site and As(2) site, respectively. The red curves are NQR spectra of the higher-quality sample (New), and the black lines are those of the early-stage sample (Previous). (d) Temperature dependence of ac susceptibility measurements at zero field. The red arrow indicates the SC transition temperature $T\subm{SC}$ estimated from the onset of the SC diamagnetic signal.}
\end{figure*}
\par
The improved homogeneity of the sample resulted in a larger internal magnetic field in the AFM state below $T\subm{SC}$. Figures 2(a) and 2(b) show the temperature variation of the NQR spectra for the As(2) and As(1) sites, respectively. While the NQR spectrum at the As(2) site became broad below $T\subm{N}$ in the early-stage sample \cite{Kibune}, the higher-quality sample exhibited a clear splitting of the NQR spectrum, suggesting a more homogeneous internal field. The splitting of the spectrum was observed below 0.3 K, which is higher than $T\subm{N} \sim 0.25$ K in the early-stage sample. The splitting width was approximately 350 kHz, corresponding to an internal field of $\mu_{0}H\subm{int}\sim$ 25 mT. This value is larger than $\mu_{0}H\subm{int} \sim 16$ mT in the early-stage sample  \cite{Ogata_c_old}, suggesting a larger AFM moment in the higher-quality sample. Higher $T\subm{N}$ and larger AFM moment in the higher-quality sample indicate that the AFM state in \CRA is intrinsic and is sensitive to disorder. Regarding the As(1) site, while almost no increase in linewidth was observed below $T\subm{N}$ in the early-stage sample \cite{Kibune}, the higher-quality sample exhibits a linewidth broadening below $T\subm{N}$. The spectral shape at the As(1) site suggests that the internal-field distribution spreads around zero field. Modifications of the magnetic structure that we proposed in our previous study based on the cancellation of the internal field at the As(1) site \cite{Ogata_ab_old} are required. As shown in Fig.~S1(b), the linewidth of the As(1) NQR spectrum in the normal state for the early-stage sample is comparable to that in the AFM state for the higher-quality sample \cite{SM}. Owing to the smaller AFM moment in the early-stage sample, it is considered that the relatively broader NQR spectrum would mask the linewidth broadening by the internal field in the AFM state at the As(1) site. This highlights the importance of results observed in the higher-quality samples.\par
While clear spectral changes were observed in the AFM state, no changes in spectral shape were observed at either As site across $T_{0} \sim 0.5$ K except for a decrease in the NQR intensity. Figure 2(c) shows the temperature evolutions of the NQR intensity multiplied by temperature $I(T)T$ of each As site. In conventional materials, $I(T)T$ remains constant, whereas it decreases when the nuclear spin relaxation time $T_2$ gets shorter or when the penetration depth of the RF pulse is reduced due to SC shielding effect. $I(T)T$ in \CRA started to decrease at $T_{0} \sim 0.5$ K \cite{QDW_new,QDW_Khanenko} and $T_2$ gets shorter below $T_0$, suggesting the enhancement of magnetic fluctuations. Decrease in $I(T)T$ below $T_0$ was also observed in the early-stage sample \cite{Kibune}. In the present study, the reduction in $I(T)T$ is more pronounced than in the early-stage sample, which is consistent with the more stronger anomaly at $T_0$ \cite{QDW_new,QDW_Khanenko}.\par
The internal field of the AFM order in \CRA exhibits a first-order-like phase transition. Figure 2(d) shows the temperature evolutions of the internal field at each As site, estimated from the NQR spectrum. $H\subm{int}$ is estimated from the peak position of the split spectrum at the As(2) site, and from the change in the FWHM of the NQR spectrum at the As(1) site. The internal field in the AFM state increases abruptly at $T\subm{N}$, while it shows only small temperature dependence below $T\subm{N}$. A critical exponent $\beta$ derived by fitting of the relation $H\subm{int}(T) = H\subm{int}(0)(1-T/T\subm{N})^{\beta}$ would be 0.025, as shown with the red dotted curve in Fig.~2(d), which is substantially lower than the conventional mean-field value of 0.5. Furthermore, as shown in Fig.~2(a), the As(2) NQR spectrum at 0.3 K exhibited the coexistence of the signals above and below the AFM transition, suggesting the presence of a phase separation. These behaviors seem to be characteristics of first-order phase transitions. However, hysteresis associated with the first-order phase transition could not be observed in the temperature steps of this NQR measurement. In the early-stage sample, the first-order-like phase transition would have been smeared out by the inhomogeneity of the AFM transition, making it difficult to detect. In the NQR measurements, the RF pulse instantaneously heats the electronic system, which then undergoes immediate thermal relaxation. Consequently, the temperature displayed by the thermometer differs from the temperature of the electronic system. Estimating the temperature of the electronic system at 0.3 K using the phase-detection method yields 0.35 K$<T<T\subm{SC}$ (see Figs.~S2(a) and S2(b) \cite{SM}). Therefore, it is suggested that $T\subm{N}$ is almost identical to $T\subm{SC}$. The lower $T\subm{N}$ compared to $T\subm{SC}$ observed in the early-stage sample may have been caused by inhomogeneity of the sample, which weakened the AFM state.\par
\begin{figure*}[htbp!]
   \begin{center}
   \includegraphics[width=0.8\linewidth]{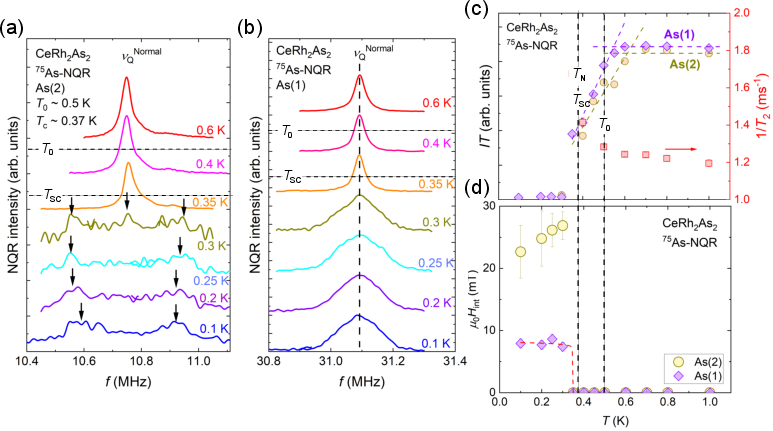}
   \end{center}
   \caption{The temperature evolutions of the $^{75}$As-NQR spectrum of (a)the As(2) and (b)As(1) sites in CeRh$_2$As$_2$. (c) The temperature evolution of the spectrum intensity ($I$) times temperature ($T$) (left axis) and nuclear spin-spin relaxation rate $1/T_2$ at the As(2) site (right axis). (d) The temperature evolutions of estimated internal field $H\subm{int}$ at the each As site. $H\subm{int}$ at the As(2) site is estimated from spectrum splitting ($H\subm{int} = \frac{1}{2}(f_{+} - f_{-})/\gamma\subm{As}$), and $H\subm{int}$ at the As(1) site is estimated from change in FWHM ($H\subm{int} = \frac{1}{2}(\mathrm{FWHM}(T)-\mathrm{FWHM}(\mathrm{normal}))/\gamma\subm{As}$).}
\end{figure*}
\par
%
To clarify the internal fields at each As site and to investigate the magnetic structure in the AFM state, we performed NMR measurements for $H\parallel c$ and $H\parallel [110]$ . As shown in Fig.~3(a), the NMR spectrum at the As(2) site splits in the AFM state by approximately 370 kHz in $H\parallel c$. In $H\parallel [110]$, the linewidth did not change significantly, but instead the peak shifts to the low-frequency side by about 60 kHz. This shift can be attributed to the internal field along the $c$ axis, which results in a tilting of the effective field at the As(2) site ($\bm{H}\subm{eff} = \bm{H}\subm{ext}+\bm{H}\subm{int}$) from [110] to the $c$ axis. The shift gives an internal field of 25 mT along the $c$-axis, which is the same value as estimated from the splitting of the NQR spectrum. The reduction of the Knight shift due to superconductivity is about one order of magnitude smaller than this shift. Similar behaviors have also been reported in the early-stage sample \cite{Ogata_ab_old}.\par
Regarding the As(1) site, the linewidth broadened by approximately 35 kHz in the $c$-axis field and by approximately 200 kHz in the [110] direction field. Since the increase in NMR linewidth due to the internal field becomes maximum when the applied field is parallel to the internal field and minimum when perpendicular, these results indicate that the internal field is mainly oriented along the $c$-axis direction for the As(2) site and within the $ab$ plane direction for the As(1) site.\par
\begin{figure*}[htbp!]
   \begin{center}
   \includegraphics[width=0.9\linewidth]{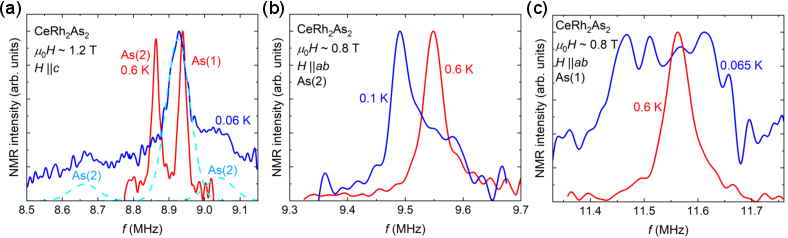}
   \end{center}
   \caption{The NMR spectra at 0.6 K (normal state) and lowest temperature (SC and AFM state) under (a) the $c$-axis field 1.2 T and (b)(c) [110] direction field. The dotted curves is the results of Gaussian fitting.}
\end{figure*}
\par
Based on above results, we analyze possible magnetic structures in the AFM state. As shown in Fig.~4(a), the NQR spectrum splits and exhibits a flat-top shape with a broadening factor in a commensurate AFM structure. However, the NQR spectrum at the As(1) site did not show such broadening. With a one-dimensional incommensurate field distribution [$H\subm{int}\propto \cos{(\bm{q}\cdot \bm{r})}$], the splitting is of the double-horn type, as shown in Fig.~4(b), which is also inconsistent with the NQR spectrum at the As (1) sites. With a two-dimensional incommensurate field distribution [$H\subm{int}\propto \cos{(\bm{q}_{1}\cdot \bm{r})}\cdot \cos{(\bm{q}_{2} \cdot \bm{r})}$], the weight of the zero internal field increases, showing a broadened single peak in Fig.~4(c). The NQR spectrum at the As(1) site in the AFM state closely resembles this shape, suggesting the magnetic structure in the AFM state should contain a two-dimensional incommensurate component. On the other hand, NQR spectrum at the As(2) site shows clear splitting. Therefore, the most plausible magnetic structure is a combination of a $\bm{Q} = \bm{0}$ A-type AFM structure with magnetic moment along the $c$ axis \cite{Ogata_ab_old} and an in-plane two-dimentional incommensurate modulation. We propose a magnetic structure shown in Fig.~4(f) as a plausible candidate. (See the supplemental material about the NQR spectrum simulation for other magnetic structures \cite{SM}). Figures~4(d) and 4(e) show the NQR-spectrum simulations based on the dipole magnetic field in this magnetic structure. This magnetic structure consistently reproduces the NMR spectra in Figs.~3(a)-(c) (see Figs. S4(a)-(c) \cite{SM}). Here, we calculated with the propagation vectors $\bm{q}_1 = [0.07 \pi, 0, 0]$, $\bm{q}_2 = [0, 0.07\pi, 0]$ and with an in-plane component half the size of the $c$-axis moment. Note that the change in propagation vector $\bm{q}$ has little effect on the calculation results and cannot be estimated from the NQR spectrum. However, incommensurate modulations near $\bm{q}\sim[\pi,\pi,0]$, as reported by ARPES and neutron scattering measurement \cite{pipi_ARPES,pipi_neutron}, cannot reproduce observed NMR spectrum because the internal field at the As(1) site has a larger $c$-axis component in such a $q$-vector (see supplemental material \cite{SM}).\par
\begin{figure}[htbp!]
   \begin{center}
   \includegraphics[width=0.95\linewidth]{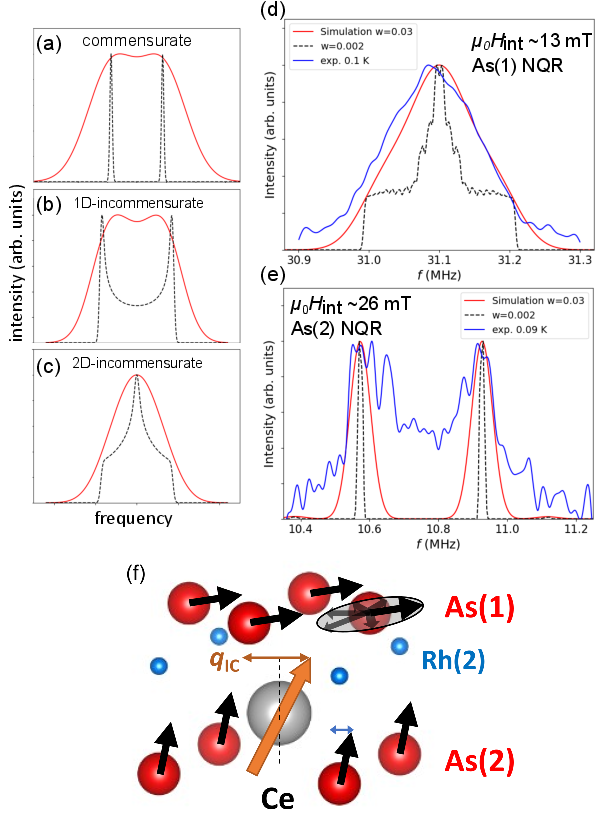}
   \end{center}
   \caption{Possible NQR spectrum with (a) commensurate, (b) 1D-incommensurate ($H\subm{int}\propto \cos{(\bm{q}\cdot \bm{r})}$), and (c) 2D-incommensurate ($H\subm{int}\propto \cos{(\bm{q}_{1}\cdot \bm{r})}\cdot \cos{(\bm{q}_{2} \cdot \bm{r})}$) internal field distribution. Black dotted curves and red solid curves represent the simulated spectra with small and large broadening factor, respectively. (d)(e) Simulated NQR spectrum in the AFM state at the As(1) site and the As(2) site, respectively. The magnetic structure we used for the simulation is in-plane two-dimensional incommensurate modulation with the magnetic structure illustrated in Fig.~1(a). Blue solid curves represent the experimental results in the lowest temperature. Black dotted curves and red solid curves represent the simulated spectra with small ($w=0.002$ MHz) and large ($w=0.03$ MHz) broadening factor, respectively. (f) An illustration depicting the magnetic structure in CeRh$_2$As$_2$. The black arrows represent the internal field at the As sites. The red arrows illustrate the tilting of the magnetic moment due to the incommensurate component. }
\end{figure}
%
In UPt$_2$Si$_2$, which also have the same CaBe$_2$Ge$_2$-type crystal structure as CeRh$_2$As$_2$, it has been reported that the magnetic structure exhibits a $\bm{Q}=\bm{0}$ A-type AFM order and an incommensurate magnetic modulation \cite{UPt2Si2_RXS,UPt2Si2_neutron}. The incommensurate $\bm{q}$ vector has the same propagation vector as that of a charge density wave state. The incommensurate modulation in \CRA is related to the order parameter in the phase I and might originate from the locally noncentrosymmetric crystal structure.\par
We discuss the correlation between phase I below $T_0 \sim 0.5$ K and the AFM state below $T\subm{N}\sim 0.35$ K. In the NQR measurements, although the internal field was observed below $T\subm{N}$, the decrease in $IT$ starts below $T_0$. This suggests the development of magnetic fluctuations. On the other hand, from the $\mu$SR measurement, it was reported that the internal field emerges at $T_0$ \cite{muSR}. We attribute this difference to the different timescales of the probe used in the measurements. In addition to the smaller gyromagnetic ratio of As nucleus $\gamma\subm{As}/2\pi =7.29$ MHz/T than that of muon $\gamma_{\mu}/2\pi =135.5$ MHz/T, there is a dead time of tens of microseconds due to the RF pulse sequence in the NQR measurements. Therefore, $\mu$SR observes quasistatic yet fluctuating moments, while NQR or magnetization measurements do not detect them \cite{QDW,QDW_Khanenko}. A slowing down of the magnetic fluctuations due to the superconductivity would allow us to detect internal field below $T\subm{N}$ as a freezing temperature. Thus, the first-order-like phase transition at $T\subm{N}$ can be understood as the fluctuation freezing at $T\subm{N}$. The close proximity of $T\subm{N}$ to $T\subm{SC}$ suggest a novel correlation between superconductivity and magnetism: the SC transition could induce the freezing of the magnetic fluctuations. The micro-Hall probe measurements also reported that the internal field emerges at $T_0$, interpreted as incomplete compensation of the internal field in the AFM domain walls \cite{microHall}. In the scenario of the fluctuation freezing, this result is attributed to the magnetic freezing at the domain walls.\par
Finally, we point out the similarity between the phase I in \CRA and the A phase in CeCu$_2$Si$_2$ \cite{CeCu2Si2_NQR,CeCu2Si2_muSR}. In the A phase of CeCu$_2$Si$_2$, the intensity of the NQR signal decreased \cite{CeCu2Si2_NQR}, while the fast decay showing the magnetism was observed in the $\mu$SR measurements \cite{CeCu2Si2_muSR}. This is similar to the magnetic state observe in $T\subm{N}<T<T_0$ in CeRh$_2$As$_2$. However, the correlation with superconductivity is different: the static magnetic order in CeCu$_2$Si$_2$ competes with superconductivity \cite{CeCu2Si2_NQR,CeCu2Si2_muSR}, whereas the phase I in \CRA not only coexists with superconductivity but seems to become a static AFM order only when superconductivity sets in. As mentioned above, the relationship between magnetism and superconductivity in \CRA is quite unusual, which deserves more investigation.\par
%
In conclusion, we performed $^{75}$As-NQR and NMR measurements on \CRA higher-quality single crystal with the locally noncentrosymmetric structure to investigate the magnetic properties. The NQR spectrum at the As(2) site exhibited a clear split below $T\subm{N}$ and the internal field in the AFM state became 1.5 times larger than in the early-stage sample. $T\subm{N}$ was higher than in the early-stage sample ($T\subm{N}\sim0.25$ K) and nearly identical to $T\subm{SC} \sim 0.37$ K. A linewidth broadening was observed even at the As(1) site where the internal field was not detected in the early-stage sample, which suggests the presence of two-dimensional incommensurate modulation in addition to the magnetic structure that we previously proposed \cite{Ogata_ab_old}. In agreement with previous results, this study did not provide evidence for a static AFM order in the temperature range $T\subm{N}\approx T\subm{SC} < T <T_0$, but they confirmed a pronounced decrease in the NQR intensity setting in at $T_0$. This suggests the onset of a fluctuating ordered magnetic state at $T_0$. At $T\subm{N}$, the abrupt, step like onset of an internal field and evidence for phase separation just around $T\subm{N}$ indicate a first-order-like onset of a static AFM order. The observation that $T\subm{N}$ is almost identical to $T\subm{SC}$ in the higher-quality samples suggests that it is the onset of superconductivity which suppresses magnetic fluctuations and induces the static order. This is a rather unique result among strongly correlated systems. Our results provide new insights into the magnetism in \CRA and provide a new key to elucidate the physics arising from the cooperation between magnetism and superconductivity, phase I, and the broken local inversion symmetry.\par
\vskip.5\baselineskip
\begin{acknowledgments}
This work was partially supported by the Kyoto University LTM Center and Grants-in-Aid for Scientific Research (KAKENHI) (Grants No. JP20KK0061, No. JP20H00130, No. JP21K18600, No. JP22H04933, No. JP22H01168, No. JP23K22439, No. JP23H01124, No. JP23K25821, No. JP24KJ1360, and No. JP25H00609). This work was also supported by JST SPRING (grant number JPMJSP2110) and research support funding from the Kyoto University Foundation, and ISHIZUE 2024 of Kyoto University Research Department Program, and Murata Science and Education Foundation. C. G. and E. H. acknowledge support from the DFG program Fermi-NESt through Grant No. GE 602/4-1. Additionally, E. H. acknowledges funding by the DFG through CRC1143 (Project No. 247310070) and the Würzburg-Dresden Cluster of Excellence on Complexity and Topology in Quantum Matter—ct.qmat (EXC 2147, Project ID 390858490). Seunghyun Khim acknowledges support from the DFG through KH 387/1-1.
\end{acknowledgments}

\bibliographystyle{apsrev4-2}
\bibliography{thesis_newNQR}

\clearpage
\onecolumngrid

\begin{center}
{\large\bfseries Supplemental Material}\\[0.5em]
{\large\bfseries Incommensurate modulation with $\bm{Q}=\bm{0}$ A-type Antiferromagnetic Order in \CRA revealed by NQR studies}\\[1em]

Shiki Ogata$^{1,*}$, Shunsaku Kitagawa$^{1}$, Kenji Ishida$^{1}$, Manuel Brando$^{2}$,\\ Elena Hassinger$^{3,2}$, Christoph Geibel$^{2}$, and Seunghyun Khim$^{2}$ 

\vspace{0.8em} 

{\small\itshape $^{1}$Department of Physics, Kyoto University, Kyoto 606-8502, Japan\\ $^{2}$Max Planck Institute for Chemical Physics of Solids, D-01187 Dresden, Germany\\ $^{3}$Institute for Quantum Materials and Technology, Karlsruhe Institute of Technology, 76131 Karlsruhe, Germany}
\end{center}

\vspace{1em}

\twocolumngrid

\setcounter{section}{0}
\setcounter{equation}{0}
\setcounter{figure}{0}
\setcounter{table}{0}

\setlength\abovecaptionskip{0pt}


\renewcommand{\thefigure}{S\arabic{figure}}
\makeatletter 
\def\@startsection#1#2#3#4#5#6{%
  \if@noskipsec \leavevmode \fi
  \par \@tempskipa #4\relax
  \@afterindenttrue
  \ifdim \@tempskipa <\z@ \@afterindentfalse\fi
  \if@nobreak
    \everypar{}%
  \else
    \addpenalty\@secpenalty\addvspace\@tempskipa
  \fi
  \@ifstar
    {\@ssect{#3}{#4}{#5}{#6}}%
    {\@dblarg{\@mysect{#1}{#2}{#3}{#4}{#5}{#6}}}}
\def\@mysect#1#2#3#4#5#6[#7]#8{%
  \ifnum #2>\c@secnumdepth \let\@svsec\@empty
  \else
    \refstepcounter{#1}%
    \protected@edef\@svsec{\csname the#1\endcsname\hskip 1em}%
  \fi
  \@tempskipa #5\relax
  \ifdim \@tempskipa>\z@
    \begingroup
      #6{%
        \@hangfrom{\hskip #3\relax\@svsec}%
        #8\par
      }%
    \endgroup
  \else
    \def\@svsechd{#6{\hskip #3\relax\@svsec #8}\hskip #5\relax}%
  \fi
  \@xsect{#5}}
\makeatother


\author{Shiki \surname{Ogata}}
\email{ogata.s.1029@m.isct.ac.jp}
\affiliation{Department of Physics, Kyoto University, Kyoto 606-8502, Japan}
\author{Shunsaku \surname{Kitagawa}}
\affiliation{Department of Physics, Kyoto University, Kyoto 606-8502, Japan}
\author{Kenji \surname{Ishida}}
\affiliation{Department of Physics, Kyoto University, Kyoto 606-8502, Japan}

\author{Manuel \surname{Brando}}
\affiliation{Max Planck Institute for Chemical Physics of Solids, D-01187 Dresden, Germany}

\author{Elena \surname{Hassinger}}
\affiliation{Technical University Dresden, Institute for Solid State and Materials Physics, 01062 Dresden, Germany}
\affiliation{Max Planck Institute for Chemical Physics of Solids, D-01187 Dresden, Germany}

\author{Christoph \surname{Geibel}}
\affiliation{Max Planck Institute for Chemical Physics of Solids, D-01187 Dresden, Germany}
\author{Seunghyun \surname{Khim}}
\affiliation{Max Planck Institute for Chemical Physics of Solids, D-01187 Dresden, Germany}
\date{\today}

\maketitle
\section{NQR spectra of the early-stage sample}
Figure S1 shows a comparison of NQR spectra between the early-stage sample (Previous) and the higher-quality sample (New) \cite{Kibune}. As shown in Fig.~S1(a), the NQR spectrum at the As(2) site in the early-stage sample was weakened due to broadening in the AFM state. On the other hand, the linewidth of the NQR spectrum at the As(2) site in the early-stage sample was slightly broader in the AFM state than in the normal state, but this broadening was less pronounced than that at the As(2) site. We interpreted it as an imperfect cancellation of the internal field at the As(1) site. However, our results about the higher-quality sample suggest that two dimentional incommensurate modulation exists and was masked by the broad NQR spectrum originating from disorder in the early-stage sample. Indeed, the linewidth in the AFM state of the early-stage sample is almost the same as that of the higher-quality sample.\par

\begin{figure*}[htbp!]
   \begin{center}
   \includegraphics[width=0.7\linewidth]{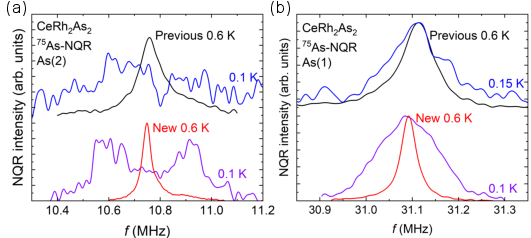}
   \end{center}
   \caption{Comparison of NQR spectra between the early-stage sample (Previous) \cite{Kibune} and the higher-quality sample (New) at the (a) As(2) site and (b) As(1) site, respectively.}
\end{figure*}
\section{Heat-up test using phase-detection method}
For an accurate estimation of $T\subm{N}$, we evaluated the heat-up effect caused by the RF pulse in the NQR measurements. The surface of the sample where the RF field penetrates is instantaneously heated up and then the heat is immediately transferred to the entire sample and $^3$He-$^4$He mixture. Consequently, the heat-up effect occurring at the moment the NQR signal is observed cannot be measured by a thermometer \cite{Heat-up_SRO}. Thus, we estimated the actual sample temperature using the phase detection method with the RF pulse. Figure~S2(a) shows a schematic of the pulse sequence for the phase-detection method \cite{Heat-up_SRO,UTe2_Knight_shift}. In this method, we apply a sufficiently weak RF pulse (detection pulse) to the NQR tank circuit with the sample and observe the returned detection pulse with the NMR receiver. The changes in the intensity or phase of the returned detection pulse related to a change in the SC diamagnetism of the sample, allowing us to examine the temperature of the electron system immediately after the NQR pulse. Figures~S2(b) and S2(c) show results of the phase-detection method. As seen in Fig.~S2(b), without NQR pulses, the intensity of the detection pulse changes below $T\subm{SC}$. When NQR pulses are applied, almost no change in the detection pulse intensity was observed at 0.1 K, which is well below $T\subm{SC}$. However, at 0.3 K, close to $T\subm{SC}$, the detection pulse intensity changed due to heating with NQR pulse powers over -14 dB. With the -12 dB NQR pulse, the intensity of the detection pulse was almost the same as that in the normal state, but with -13 dB NQR pulse, it was lower (0.35 K $< T < T\subm{SC}$). Figure~S2(d) shows the change in the NQR spectrum at the As(2) site at 0.3 K caused by heating. With -12 dB NQR pulse, where the system is in the normal state, only a peak at the normal-state position was observed. In contrast, with -13 dB NQR pulse (0.35 K $< T < T\subm{SC}$), both the AFM and normal peaks coexist, indicating $T\subm{N}$ is quite close to $T\subm{SC}$. The coexistence of the AFM and normal peaks suggests phase separation. The lower $T\subm{N}$ compared to $T\subm{SC}$ observed in the early-stage sample may have been caused by inhomogeneity within the sample which weakened the AFM state.\par

\begin{figure*}[htbp!]
   \begin{center}
   \includegraphics[width=0.8\linewidth]{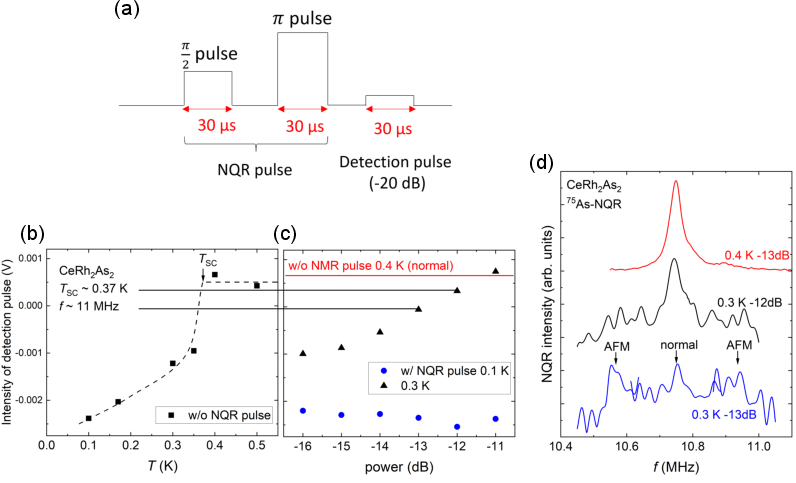}
   \end{center}
   \caption{(a) The schematic of the pulse sequence for the phase-detection method. (b)Temperature evolution and (c) NQR pulse power dependence of the intensity of the detection pulse. The dotted curve is guide to the eye. (d) The NQR spectra at the As(2) site at 0.3 K with deferent NQR pulse power. The NQR spectrum at 0.4 K is also shown for reference.}
\end{figure*}
\par
\section{NQR/NMR spectrum simulation with other magnetic structures}
We simulated NQR spectra for other magnetic structures, which are not mentioned in the main text. We simulated with the one-dimensional incommensurate in-plane moment (Figs.~S3(a) and S3(b)), the two-dimensional incommensurate $c$-axis moment (Figs.~S3(c) and S3(d)), and the two-dimensional incommensurate in-plane moment with a propagation vector $\bm{q}$ close to $\pi$ (Figs.~S3(e) and S3(f)), in addition to the $\bm{Q}=\bm{0}$ A-type AFM structure. Except for the third magnetic structure, they do not reproduce the NQR spectra at the As(1) site. Even the third magnetic structure does not reproduce the NMR spectrum under $H\parallel ab$ (Fig. S3(h)), because the main component of the internal field at the As(1) site is along the $c$ axis.\par
\begin{figure*}[htbp!]
   \begin{center}
   \includegraphics[width=0.6\linewidth]{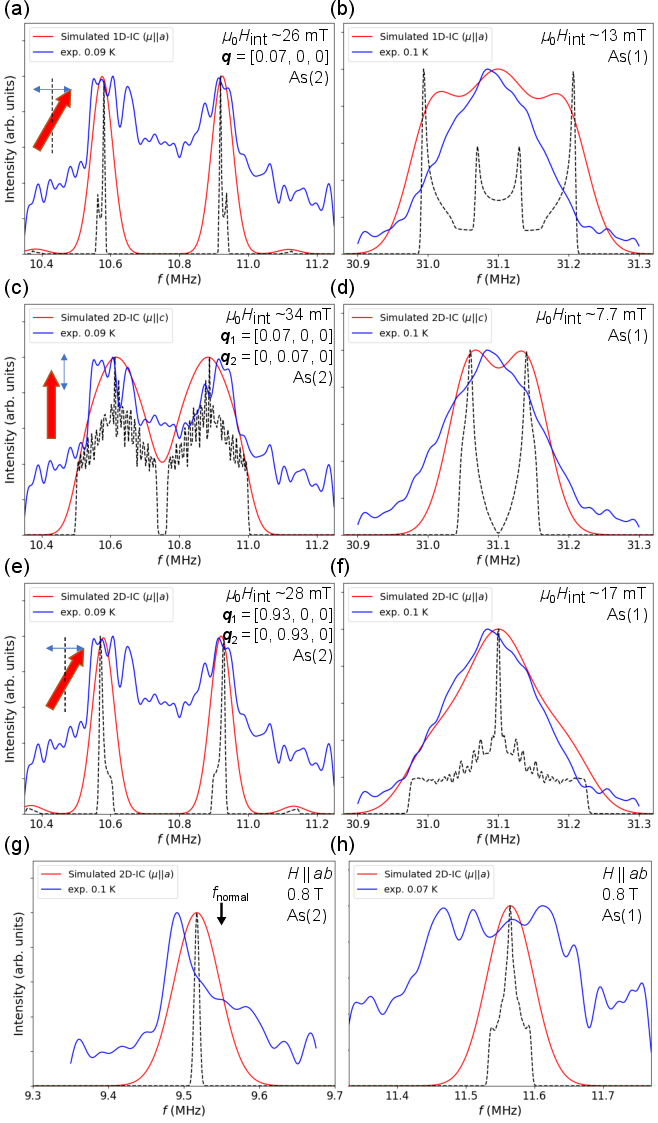}
   \end{center}
   \caption{The simulated NQR spectra with other magnetic structures. The magnetic structures are a $\bm{Q}=\bm{0}$ A-type AFM structure plus (a)(b) the one-dimensional incommensurate in-plane moment, (c)(d) the two-dimensional incommensurate $c$-moment, and (e)(f) the two-dimensional incommensurate in-plane moment with a propagation vector $\bm{q}$ close to $\pi$. The red arrows represent the magnetic structures. (g)(f) The simulated NMR spectra in the $ab$-plane field 0.8 T with the magnetic structure we used in (e)(f).}
\end{figure*}
\newpage
\section{NMR spectrum with 2D-incommensurate modulation}
Figure S4 shows a comparison between NMR spectrum simulations and experimental results in the AFM state. The magnetic structure is two-dimensional incommensurate modulation ($\bm{\mu}\parallel ab$) in addition to the $\bm{Q}=\bm{0}$ A-type AFM state ($\bm{\mu}\parallel c$). The parameters for the magnetic structure are the same as those used in the NQR spectrum simulations in the main text. As shown in Fig.~S4(a), the spectrum simulation reproduces the split peak of the As(2) site well for the $c$-axis field of 1.2 T. For the [110] field, it reproduces the shift toward the low-frequency side due to the internal field in the AFM state at As(2) site, as shown in Fig.~S4(b), although the reduction is slightly smaller in simulation. Such a negative shift was also observed in the early-stage sample \cite{Ogata_ab_old}. For the As(2) site, the simulation reproduces the linewidth broadening.
\begin{figure*}[htbp!]
   \begin{center}
   \includegraphics[width=1\linewidth]{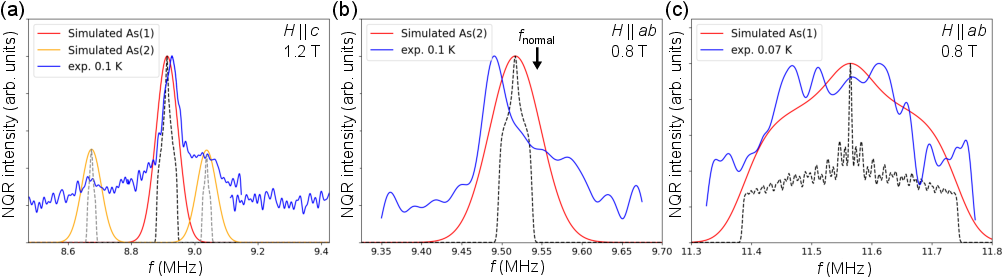}
   \end{center}
   \caption{The NMR spectrum simulation (a) in the $c$-axis field and (b)(c) the [110] field assuming an in-plane two-dimensional incommensurate modulation with a $\bm{Q}=\bm{0}$ A-type AFM structure. The red and orange curves indicate the simulated spectra at the As sites, and the blue curves indicate the experimental results in the lowest temperatures. The dotted curves indicate the simulated spectra with small broadening factor. The arrow in (b) represents the peak frequency in the normal state.}
\end{figure*}
\par
%



\end{document}